\documentclass[twocolumn,prl,amsmath,amssymb,superscriptaddress]{revtex4-1}
\usepackage[utf8]{inputenc}
\usepackage[english]{babel}
\usepackage{amsfonts,amsmath,amssymb}
\usepackage{graphicx}
\usepackage{bm} 
\usepackage{mathrsfs}
\usepackage{subfigure}
\usepackage{textcomp}
\usepackage{hyperref}
\usepackage[flushleft]{threeparttable}
\usepackage[per-mode=symbol]{siunitx}
\usepackage[version=3]{mhchem}

\DeclareSIUnit\intensity{\watt\per\centi\meter\squared}
\DeclareSIUnit\fieldstrength{\volt\per\centi\meter}

\def\qr{{\bf r}}
\def\qR{{\bf R}}

\newcommand{\degree}{\ensuremath{^\circ}}%
\newcommand{\cost}{\ensuremath{\langle\cos^2\theta_\text{2D}\rangle}}

\newcommand{\ran}{\rangle}
\newcommand{\lan}{\langle}

\newlength{\figwidth}
\newlength{\figwidthsmall}
\let\orgautoref\autoref
\providecommand{\Autoref}{%
  \def\equationautorefname{Equation}%
  \def\figureautorefname{Figure}%
  \def\subfigureautorefname{Figure}%
  \orgautoref}
 \renewcommand{\autoref}{%
  \def\equationautorefname{Eq.}%
  \def\figureautorefname{Fig.}%
  \def\subfigureautorefname{Fig.}%
  \orgautoref}

 \usepackage{color}
\usepackage[normalem]{ulem}
\definecolor{darkgreen}{rgb}{0.0,0.7,0.0}

\newcommand{\be}{\hat b}
\newcommand{\bed}{\hat b^\dagger}
\newcommand{\bra}[1]{\langle #1 | \,}
\newcommand{\ket}[1]{\, | #1 \rangle}
\newcommand{\veck}{\mathbf k}
 
\begin{document}

\title{Laser-induced rotation of iodine molecules in He-nanodroplets:\\ revivals and breaking-free}

\author{Benjamin Shepperson}
\affiliation{Department of Chemistry, Aarhus University, 8000 Aarhus C, Denmark}

\author{Anders A. S{\o}ndergaard}
\affiliation{Department of Physics and Astronomy, Aarhus University, 8000 Aarhus C, Denmark}

\author{Lars Christiansen}
\affiliation{Department of Chemistry, Aarhus University, 8000 Aarhus C, Denmark}

\author{Jan Kaczmarczyk}
\affiliation{IST Austria (Institute of Science and Technology Austria), Am Campus 1, 3400 Klosterneuburg, Austria}

\author{Robert E. Zillich}
\email[Corresponding author: ]{Robert.Zillich@jku.at}%
\affiliation{Institute for Theoretical Physics, Johannes Kepler Universit\"{a}t Linz, Altenbergerstraße 69, A-4040 Linz, Austria}

\author{Mikhail Lemeshko}
\email[Corresponding author: ]{mikhail.lemeshko@ist.ac.at}%
\affiliation{IST Austria (Institute of Science and Technology Austria), Am Campus 1, 3400 Klosterneuburg, Austria}

\author{Henrik Stapelfeldt}%
\email[Corresponding author: ]{henriks@chem.au.dk}%
\affiliation{Department of Chemistry, Aarhus University, 8000 Aarhus C, Denmark}

\date{\today}

\begin{abstract}
Rotation of molecules embedded in He nanodroplets is explored by a combination of fs laser-induced alignment experiments and angulon quasiparticle theory. We demonstrate that at low fluence of the fs alignment pulse, the molecule and its solvation shell can be set into coherent collective rotation lasting long enough to form revivals. With increasing fluence, however, the revivals disappear -- instead, rotational dynamics as rapid as for an isolated molecule is observed during the first few picoseconds. Classical calculations trace this phenomenon to transient decoupling of the molecule from its He shell. Our results open novel opportunities for studying non-equilibrium solute-solvent dynamics and quantum thermalization.
\end{abstract}

\pacs{add pacs}

\maketitle

Usually, molecules dissolved in a liquid are not rotating freely due to the intermolecular forces exerted by the surrounding solvent. An important exception is molecules embedded in liquid helium nanodroplets where high-resolution infrared~\cite{grebenev_rotational_2000} and microwave~\cite{lehnig_rotational_2009} spectroscopies display discrete rotational structure. These observations along with theoretical modelling has established a picture that molecules inside He nanodroplets can rotate frictionless although followed by a local solvation shell of He atoms. This shell increases the effective molecular moment of inertia compared to the gas-phase value~\cite{toennies_superfluid_2004,choi_infrared_2006}.

These unique properties build the expectation that it should be possible to induce frictionless rotation of molecules inside Helium droplets and follow it in real time. For isolated molecules versatile techniques based on moderately intense fs or ps laser pulses have been developed to control the rotational degrees of freedom~\cite{stapelfeldt_colloquium:_2003,ohshima_coherent_2010,korobenko_rotational_2014}. In particular, such methods have been extensively used to confine molecular axes to laboratory-fixed axes -- methods referred to as alignment and orientation~\cite{stapelfeldt_colloquium:_2003}. Recently, the first time-resolved experiments of molecular rotation inside He droplets revealed that moderately intense laser pulses can induce alignment of molecules~\cite{pentlehner_impulsive_2013,christiansen_alignment_2015}. The measurements showed, however, no sign of frictionless rotation. Notably, the transient alignment-recurrences (revivals) characteristic of freely rotating molecules in gas phase were absent. These observations seemed at odds with the prevailing conception of rotational structure obtained through spectroscopy~\cite{toennies_superfluid_2004,choi_infrared_2006}.

Here we experimentally demonstrate that a sufficiently weak fs pulse can initiate coherent rotation of iodine molecules together with their He solvation shell -- lasting long enough to form revivals. Our observations are rationalized by a quantum theory based on the angulon quasiparticle~\cite{schmidt_rotation_2015, LemSchmidtChapter, schmidt_deformation_2016, Yakaboylu17, MidyaPRA16, Li16}. For strong alignment pulses the revivals disappear and, instead, strikingly fast rotational dynamics appears immediately after the pulse. Classical estimates indicate that, in this regime, He atoms of the solvation shell detach
from the molecule due to the centrifugal force generated by the rapid rotation. This can be seen as a sudden decoupling of the molecule from its solvent and for a short time the rotational motion resembles that of a free molecule.

In our experiment, 10-nm-diameter helium droplets -- each doped with at most one iodine (\ce{I_2}) molecule -- are first irradiated by a \SI{450}{fs} linearly polarized laser pulse at \SI{800}{nm}. The purpose of this kick pulse is to induce alignment of the molecules, i.e. confine their I--I internuclear axis along the polarization direction~\cite{stapelfeldt_colloquium:_2003}. Next, the molecules are Coulomb exploded by a delayed, intense probe pulse (\SI{40}{fs}, $\SI{3.7e14}{W/cm^2}$) which produces \ce{IHe^+} ion fragments with recoil directions given by the angular distribution of the molecular axes at the instant of the probe pulse. By detecting the emission directions of the \ce{IHe^+} ions with a 2D imaging detector at many different kick-probe delays, $t$, the time-dependent degree of alignment, $\cost$, can be determined -- $\theta_\text{2D}$ being the angle between the alignment pulse polarization and the projection of an \ce{IHe^+} ion velocity vector on the detector~\cite{sondergaard-jcp-2017}. More details on the experimental setup are provided in the Supplemental Material~\cite{supplement}.

\begin{figure}
  \includegraphics[width=\columnwidth]{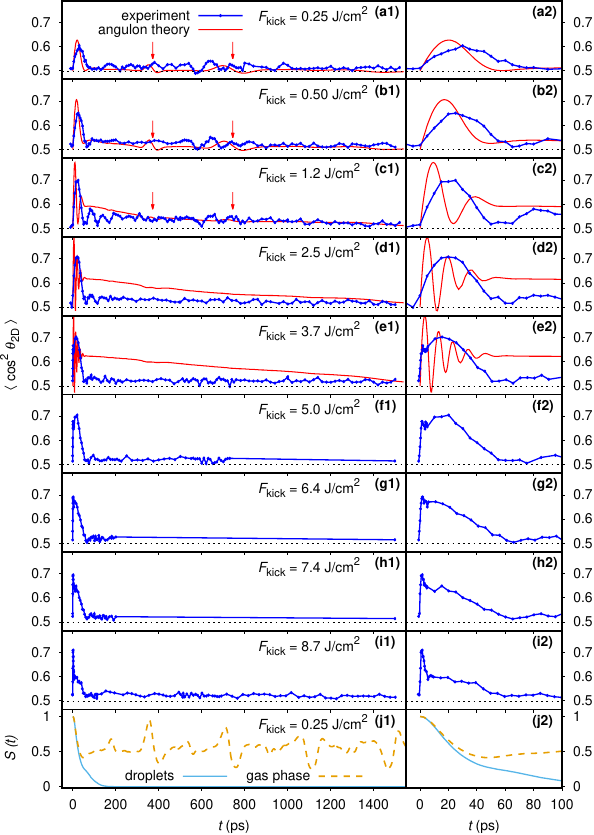}
  \caption{ The degree of alignment, $\cost$, as a function of time at different fluences of the kick pulse (centered at $t = 0$); blue curves: experimental results, red curves: results from the angulon theory. In panel (f1) the time-interval \SI{750}-\SI{1500}{ps} and in panels (g1) and (h1) the time-interval \SI{200}-\SI{1500}{ps} are shown as straight lines because, for experimental reasons, $\cost$ was not recorded in these regions.
  The right column of panels expands on the first \SI{100}{ps} to highlight the structure that starts to appear at \mbox{$F_\text{kick}$ = 3.7 J/cm$^2$}
  immediately after the kick pulse and grows to a sharp peak with maximum at \mbox{$t = \SI{1.3}{ps}$} for $F_\text{kick}$ = 8.7 J/cm$^2$. Panels (j) show the survival probability of the initial state, as defined in the text. }
\label{fig:fluences}
\end{figure}

\Autoref{fig:fluences} shows $\cost$ as a function of time for a series of
different fluences of the kick pulse, $F_\text{kick}$. At low $F_\text{kick}$
there is a distinct maximum in $\cost$ shortly after the kick pulse
[\autoref{fig:fluences}(a)-(d)]. The prompt peak grows in amplitude and appears
earlier as $F_\text{kick}$ is increased [\autoref{fig:fluences}(a2)-(d2)]. This
behavior is the result of faster rotation and more efficient alignment induced
by a stronger kick pulse and appears similar to previous measurements on
\ce{CH_3I} molecules in He droplets~\cite{pentlehner_impulsive_2013}. The
current data exhibit, however, new, previously unobserved features. First, at
$F_\text{kick}$ = 1.2 J/cm$^2$ the prompt alignment peak is followed by
pronounced yet decreasing oscillations out to $\sim\SI{200}{ps}$. Second, for
$F_\text{kick}$ = 0.25, 0.50 and 1.2 J/cm$^2$ an oscillatory structure is
observed in the interval $\SI{550}-\SI{750}{ps}$. The structure is very similar
for the three fluences with local maxima and minima at essentially the same
times. Third, on average the $\cost$ curves are gradually decaying in the range
$\sim\SI{100}-\SI{1500}{ps}$ for $F_\text{kick}$ = 0.50, 1.2 and 2.5 J/cm$^2$.

For $F_\text{kick}$ $\geq$ 2.5 J/cm$^2$ the structure in the $\SI{550}-\SI{750}{ps}$ interval disappears. Also, the oscillations after the main peak are strongly reduced for $F_\text{kick}$ = 2.5 J/cm$^2$ and essentially absent at larger fluences. Instead a substructure in the prompt alignment peak starts to appear at $F_\text{kick}$ = 3.7 J/cm$^2$ [\autoref{fig:fluences}(e2)]. As the fluence is increased the substructure grows to a prominent sharp peak ending with a maximum already at $t\sim\SI{1.3}{ps}$ for $F_\text{kick}$ = 8.7 J/cm$^2$ [Figs. \ref{fig:fluences}(i) and \ref{fig:align_molecules}].

\begin{figure}
   \includegraphics[width=\columnwidth]{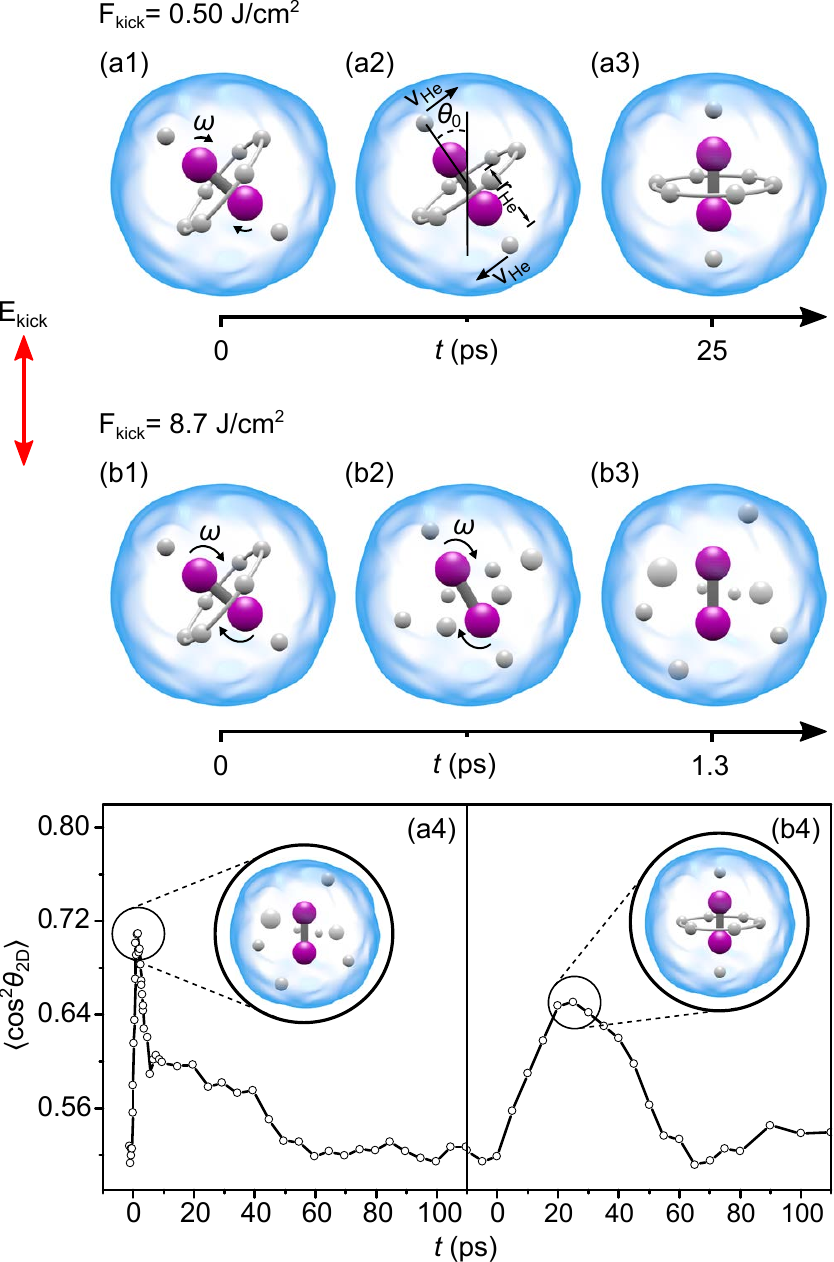}
  \caption{Schematic illustration of laser-induced rotation of \ce{I_2} molecules inside He droplets, based on
     the classical model described in the text, for a weak [(a1)-(a3)] and a
     strong [(b1)-(b3)] kick pulse. (a4) and (b4) relate the illustrations to the data recorded.
     (a2) illustrates parameters used in the classical model. $\theta_0$: The angle between the molecular axis and the kick pulse polarization just prior to the laser-molecule interaction. $r_\text{He}$: Distance from the He atom at the ends to the axis of rotation. $v_\text{He}$: The linear speed of the He atoms at the ends of the molecule gained from the laser-molecule interaction.}
  \label{fig:cartoon}
\end{figure}

\begin{figure}
  \includegraphics[width=1\columnwidth]{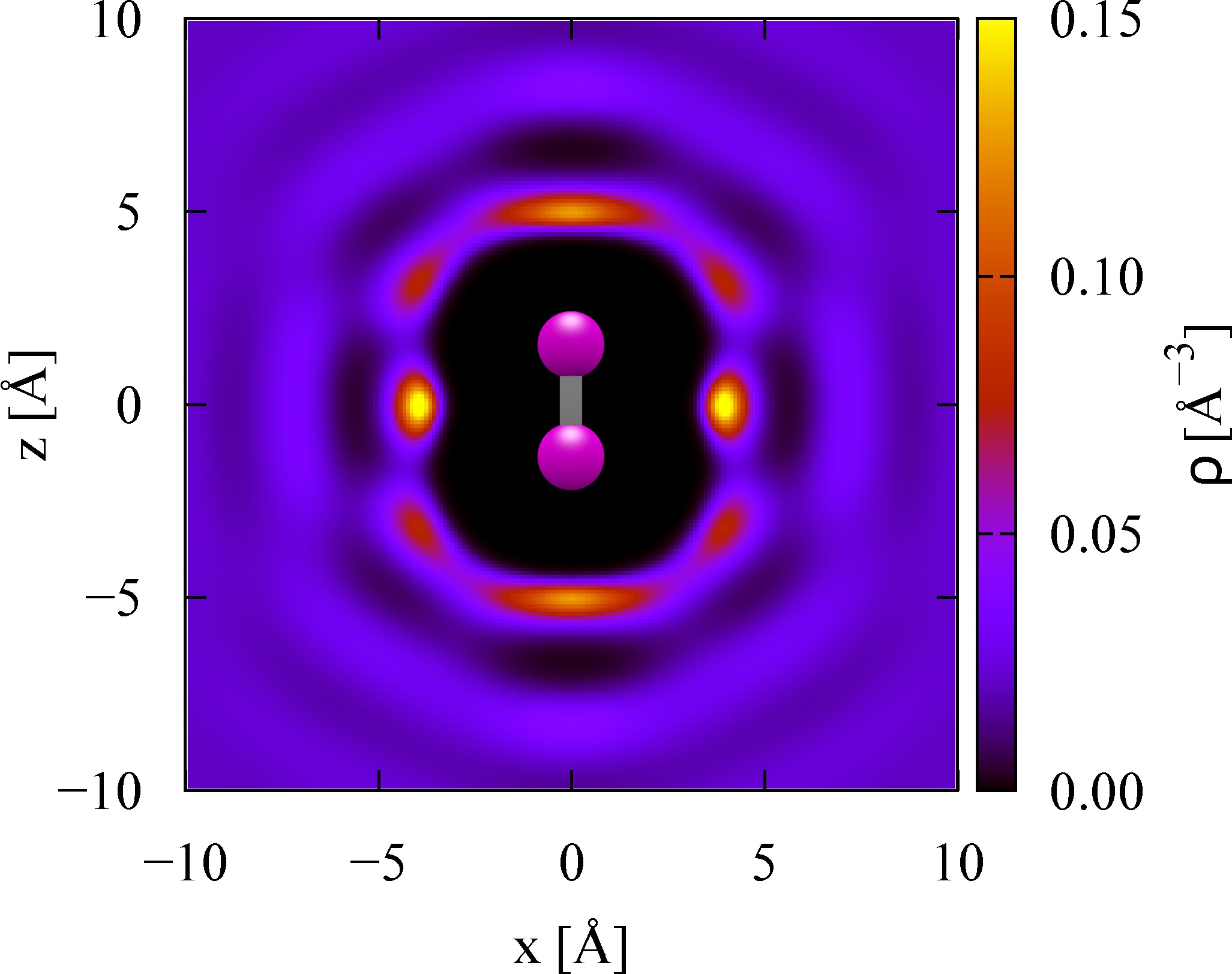}
  \caption{ He density, $\rho$, around I$_2$ in the molecular frame in equilibrium. It is obtained from a path integral Monte Carlo calculation~\cite{supplement} and corresponds to the situation prior to the kick pulse.}
  \label{fig:dens}
\end{figure}

\begin{figure}
  \includegraphics[width=\columnwidth]{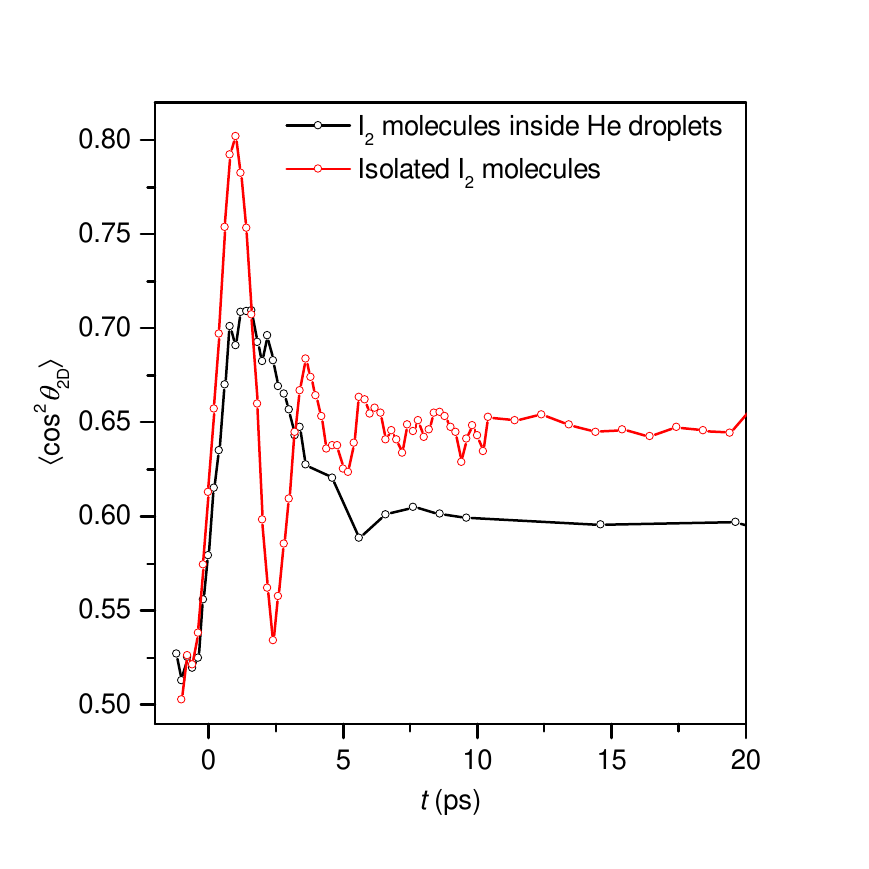}
  \caption{The degree alignment, $\cost$, at early times for isolated \ce{I_2} molecules and \ce{I_2} molecules in He droplets recorded for $F_\text{kick}$ = 8.7 J/cm$^2$. The laser parameters of both the kick pulse and the probe pulse were identical for the measurements on the isolated molecules and on the molecules in He droplets.}
  \label{fig:align_molecules}
\end{figure}

We interpret the oscillations after the prompt peak and the
$\SI{550}-\SI{750}{ps}$ structure as manifestations of coherent
rotation of the molecules and their local He solvation shell –- hereafter
termed He-dressed molecules. To substantiate this interpretation we first model
He-dressed molecules as classical rigid rotors driven by the polarizability
interaction with the kick pulse. A He-dressed molecule initially at an
angle $\theta_0$ to the kick pulse polarization [\autoref{fig:cartoon}(a3)] gains an angular velocity, $\omega$,
of~\cite{leibscher_enhanced_2004}:
\begin{equation}
\label{eq:omega} \omega = \frac{1}{2} \frac{\Delta \alpha F_\text{kick} \sin(2\theta_0)}{I_\text{eff}\varepsilon_0 c},
\end{equation}
where $\Delta\alpha$ is the polarizability anisotropy of \ce{I_2} and $I_\text{eff}$ is the effective moment of inertia of \ce{I_2} in the droplets. No experimental value exists for $I_\text{eff}$ so we determined it by a path integral Monte Carlo calculation~\cite{zillichJCP05}, which gave $I_\text{eff}=1.7\times I_0$ where $I_0$ is the moment of inertia of the bare \ce{I_2} molecule. The calculated He density around the \ce{I_2} molecule is shown in \autoref{fig:dens}. In our classical calculations a He-dressed molecule is treated as an \ce{I_2} molecule rigidly attached to eight He atoms placed in the minima of the \ce{I_2}--He potential~\cite{grebenev_rotational_2000,garcia-gutierrez_intermolecular_2009} (six He atoms in the central ring around the molecule and two at the ends), see \autoref{fig:cartoon}. The value of $I_\text{eff}$ determined from this structure (\autoref{fig:cartoon}) is essentially equal to the Monte Carlo one.

Equation \eqref{eq:omega} predicts that the He-dressed molecules
are set into end-over-end rotation (\autoref{fig:cartoon})
 leading to a prompt alignment peak as observed experimentally. Continued
rotation for extended times requires that superfluidity of the droplets is undistorted. A simple classical
criterion for this is that the linear speed of the
outer components of the He-dressed molecules should not exceed the Landau velocity, $v_L =56$~m/s~\cite{brauer_critical_2013}.
The highest linear speed is calculated as
$v_\text{He} = \omega r_\text{He}$, where $r_\text{He}$ is
the distance from a He atom at the ends to the axis of rotation (see
\autoref{fig:cartoon}). \Autoref{tab:theory} displays the values of $\omega$
and $v_\text{He}$ for the nine different fluences used in the experiment.
For $F_\text{kick}$  = 0.25 and 0.50 J/cm$^2$ $v_\text{He}$ $<$
$v_L$, whereas at $F_\text{kick}$ = 1.2 J/cm$^2$
$v_\text{He}$ is just above $v_L$. At higher fluences,
$v_\text{He} \gg v_L$ for almost all of the He-dressed
molecules -- independent of their initial orientation. These simple classical
considerations indicate that long-time coherent rotational dynamics of the
He-dressed molecules is only possible for the three lowest fluences -- in
accordance with the observations -- and illustrated by panels (a1)-(a3) of
\autoref{fig:cartoon}.

To elucidate the quantum dynamics of the system, we apply the recently-developed
angulon theory~\cite{LemeshkoDroplets16, schmidt_rotation_2015, LemSchmidtChapter, schmidt_deformation_2016, Yakaboylu17, MidyaPRA16, Li16, RedchenkoCPC16}. The angulon
represents a quasiparticle consisting of a molecular rotor dressed
by a many-body field of superfluid excitations, and can be thought of as a
quantum formulation of the He-dressed molecule.
 Recently it was shown that molecules in superfluid helium form
angulons~\cite{LemeshkoDroplets16}. The case of I$_2$ in helium belongs to
the strong-coupling regime, where the molecular kinetic energy is  small compared to the molecule-helium interactions~\cite{LemeshkoDroplets16, schmidt_deformation_2016}. In this regime, the angulon theory furnishes a closed-form expression for the alignment cosine:
\begin{multline}\label{eq:cos2}
\lan \cos^2{\hat{\theta}_{\rm 2D}}\ran (t) \\
= \sum_{j, j', j_1, j_2, m} c^*_{j'} c_{j} e^{i t (E_{j_2} + E_{j'} - E_{j_1} - E_{j}) / 2}
  \lan j_2 m| \cos^2{\hat{\theta}_{2D}} |j_1 m\ran \\
\times  \int d\Omega_1 \int d\Omega_2 \: Y^*_{j' m_0}(\Omega_2) Y_{j_2 m}(\Omega_2) Y_{j m_0}(\Omega_1) Y^*_{j_1 m}(\Omega_1) \\
\times e^{\alpha_1 t^2  \left[  \frac{4 \pi}{5}  \sum_{\mu}   Y_{2, \mu}(\Omega_2) Y^*_{2, \mu}(\Omega_1) - 1 \right]}.
\end{multline}
(in units of $\hbar \equiv 1$). Here $E_j = B_{\rm eff} j(j+1)$ are the molecular rotational energies, with $B_{\rm eff}$ the effective rotational constant of \ce{I_2}. $\alpha_1$ parametrizes the anizotropic molecule-helium interactions, with the strong-coupling regime defined by $B_{\rm eff} \ll \sqrt{\alpha_1}$. The coefficients, $c_j = \bra{j, m_0} \exp ( \eta \cos^2 \hat{\theta} ) \ket{j_0, m_0}$, describe the rotational wavepacket created from the initial molecular state $\ket{j_0, m_0}$ by a short laser pulse with  a dimensionless intensity $\eta$. In order to compare the theory to experiment, the results of Eq.~\eqref{eq:cos2} were averaged over the thermal distribution of the initial states and the finite width of rotational lines due to dephasing was accounted for. More details on the theoretical approach are provided in the Supplemental Material~\cite{supplement}.

The strong-coupling angulon theory is straightforward to apply if the molecule-laser interaction energy
$\eta \lesssim  \sqrt{\alpha_1}$. This is the case in \autoref{fig:fluences} (a) and (b), where $\eta /
\sqrt{\alpha_1} \approx 1.4$ and $2.7$, respectively. Both calculated $\cost$ curves (red) are dominated by
a prominent peak at early times. For $F_\text{kick}$  = 0.25 and 0.50 J/cm$^2$
the prompt alignment peak agrees with the experimental curves although the peak
amplitude is somewhat higher for the calculated curves. We ascribe this to an
underestimation of the measured degree of alignment due to non-axial
recoil effects in the Coulomb explosion process, caused by the He
environment~\cite{christensen_deconvoluting_2016}. The fluence of
$F_\text{kick}$  = 1.2 J/cm$^2$ corresponds to $\eta /  \sqrt{\alpha_1} \approx
7$ and therefore lies  beyond the reach of the strong-coupling angulon theory which predicts a faster initial dynamics compared to the experiment.
Nevertheless, the long-time decay of alignment observed experimentally is
reproduced. For higher fluences, $F_\text{kick}$  = 2.5 J/cm$^2$ and
$F_\text{kick}$  = 3.7 J/cm$^2$,  the
molecule-laser interactions dominate over molecule-helium interactions,
since $\eta /
\sqrt{\alpha_1} \approx 14$ and $21$, respectively.

For $F_\text{kick}$  = 0.25 and 0.50 J/cm$^2$ low-amplitude structures just before \SI{400} and \SI{800}{ps} are visible. The angulon model identifies these as the half- and full-revival of the He-dressed molecule,  marked by red arrows for the theoretical curves. The locations of the revival structures match $h/(4B_\text{eff})$ and $h/(2B_\text{eff})$, with $B_\text{eff} = \hbar^2/(2 I_\text{eff})$, similar to the well-studied case of isolated molecules. The magnitude of the revivals decreases for larger fluences, and they are no longer visible for  F$_\text{kick}\ge   2.5$ J/cm$^2$.

\begin{table}
     \begin{tabular*}{\columnwidth}{r@{\extracolsep{\fill}}rrr@{\extracolsep{0pt}}r}
          \hline
          $F_\text{kick}$, \si{\joule/\centi\meter\squared} & $\omega$, \num{e10} \si{\hertz} & $v_\text{He}$, \si{\meter/\second} & $E_\text{rot}(\text{He})$, \si{\centi\meter^{-1}} \\
          \hline
          \num{0.25} & \num{2.7} & \phantom{0}\num{13} & \num{0.029} \\
          \num{0.50} & \num{5.5} & \phantom{0}\num{26} & \num{0.12} \\
          \num{1.2} & \num{14} & \phantom{0}\num{65} & \num{0.71} \\
          \num{2.5} & \num{27} & \num{130} & \num{2.8} \\
          \num{3.7} & \num{41} & \num{195} & \num{6.4} \\
          \num{5.0} & \num{54} & \num{260} & \num{11} \\
          \num{6.4} & \num{70} & \num{338} & \num{19} \\
          \num{7.4} & \num{81} & \num{390} & \num{26} \\
          \num{8.7} & \num{95} & \num{454} & \num{35} \\
          \hline
     \end{tabular*}
   \caption{Classical calculation of the maximum angular velocity using Eq.~\eqref{eq:omega} with $\theta_0 = \SI{45}{\degree}$ for the nine different fluences used in the experiment. From $\omega$ the linear speed, $v_\text{He}$, and the rotational energy, $E_\text{rot}(\text{He})$, of the He atoms at the ends of the molecules are calculated -- see text.}
  \label{tab:theory}
\end{table}

Importantly, the angulon theory predicts that rotational revivals are possible for molecules strongly interacting with superfluid helium. Therefore, we interpret the observed oscillatory structure in the $\SI{550}-\SI{750}{ps}$ interval as a full rotational revival of the He-dressed molecule.
Furthermore, we note that the model captures the overall decay of $\cost$ observed  most clearly for $F_\text{kick}$  = 0.50 and 1.2 J/cm$^2$.
We quantify this decay by the survival probability (closely related to the Loschmidt echo~\cite{zangara_loschmidt_2012}),
  $S(t) =
\vert \bra{\psi(t)} \psi_0 \rangle \vert^2 \equiv \vert \bra{\psi_0} e^{i H t}
\ket{\psi_0} \vert^2$, of the state $\ket{\psi_0}$
immediately after the kick pulse excitation during time evolution under the angulon
Hamiltonian $H$~\cite{supplement}. The time-dependence of the survival probability is shown in
\autoref{fig:fluences}j. While the gas-phase survival probability (dashed line)
exhibits revivals similar to the gas-phase molecular alignment, for I$_2$ in helium
we  predict a Gaussian decay $S(t) \sim \exp(-\alpha_1 t^2)$~\cite{supplement}. The latter comes from redistribution of the angular momentum
between the molecule and the superfluid. Note that this decay occurs on a faster timescale compared to the exponential decay common for Markovian reservoirs~\cite{ramakrishna_intense_2005}.

In the high-fluence regime, $\eta/\sqrt{\alpha_1} \gg 1$, the strong-coupling
angulon theory is not applicable. However, classically, a high-fluence pulse can
induce such a fast rotation of the He-dressed molecule that helium atoms detach due to the centrifugal force -- a mechanism which, we believe, is responsible for the sharp structure appearing in the prompt alignment peak. A simple criterion for detaching one He atom is:
$E_\text{rot}(\text{He}) > E_\text{binding}(\text{He})$,
where $E_\text{rot}(\text{He})=\tfrac{1}{2}m_\text{He}$ $r_\text{He}^2 \omega^2$ is the rotational energy of a He atom
and $E_\text{binding}(\text{He}) \approx 16$~cm$^{-1}$ is the ground-state
binding energy of the \ce{HeI_2} complex ~\cite{garcia-gutierrez_intermolecular_2009,ray_experimental_2006}.
\Autoref{tab:theory} displays $E_\text{rot}(\text{He})$ calculated for the different
fluences. At F$_\text{kick} \gtrsim$ 6 J/cm$^2$ the criterion is met implying
that one or indeed several He atoms detach from the molecule (lower panels in \autoref{fig:cartoon}) since the binding energies of the first few He atoms are similar~\cite{paesani_interaction_2004}.

\Autoref{fig:align_molecules} compares the short-time alignment dynamics of molecules
in He droplets to that of isolated molecules at $F_\text{kick}$ = 8.7 J/cm$^2$. In droplets, $\cost$ evolves almost as
fast as $\cost$ of isolated molecules during the first $\sim$\SI{2}{ps}.
In classical terms, this indicates that \ce{I_2} rotates almost freely, detached from the He atoms. We observed the same rapid short-time alignment dynamics for \ce{OCS} and \ce{CS_2} molecules in He droplets. At t $>$ \SI{2}{ps} the free rotation
is quenched, which indicates a dynamical re-formation of the He-dressed molecule.

Our results demonstrate that for molecules embedded in He droplets a moderately intense laser pulse can induce coherent collective rotation of a molecule and its solvation shell for times long enough to form rotational revivals. These findings reconcile femtosecond laser-induced molecular alignment and high-resolution infrared and microwave spectroscopy. Our angulon quasiparticle theory
rationalizes the observations for the low-fluence experimental results. Future generalization of the theory may lead to a quantitative agreement
with experiments in a broad range of laser fluences and is expected
to provide new insights into the superfluid behavior of He droplets. Finally,
the observed decoupling of the molecule from its He-solvation shell
at high fluences draws parallels to the nonlinear response in the solute-solvent interaction of
rapidly rotating \ce{CN} molecules dissolved in ethanol~\cite{moskun_rotational_2006,tao_molecular_2006}.
Our results open unique opportunities for real-time studies of non-equilibrium solute-solvent dynamics, for instance,
by gradually modifying the solvation shell through insertion of other noble gas
atoms or even water molecules~\cite{choi_infrared_2006}. Furthermore,
experiments on molecules in small helium droplets might
yield insight into quantum thermalization of finite many-particle systems~\cite{RigolNature08, PolkovnikovRMP11}.

We thank Richard Schmidt for insightful discussions. JK acknowledges support from People Programme (Marie Curie Actions) of the European Union's Seventh Framework Programme (FP7/2007-2013) under REA grant agreement No. [291734]. REZ acknowledges support from the Austrian Science Fund (FWF) under grant No. P23535-N20. ML acknowledges support from the Austrian Science Foundation (FWF), under grant No. P29902-N27.
HS acknowledges support from the European Research Council-AdG (Project No. 320459, DropletControl) and the Villum Foundation.

%


\newpage
\clearpage

\vspace{0cm}

\renewcommand{\thepage}{S\arabic{page}}  
\renewcommand{\thesection}{S\arabic{section}}   
\renewcommand{\thetable}{S\arabic{table}}   
\renewcommand{\thefigure}{S\arabic{figure}}
\renewcommand{\theequation}{S\arabic{equation}}

\setcounter{figure}{0} 
\setcounter{table}{0} 
\setcounter{equation}{0} 
\setcounter{page}{1} 

\onecolumngrid
\section{Supplemental Material}

\appendix

\section{Experimental setup and method}

\begin{figure}[b]
  \includegraphics[width=1.3\figwidth]{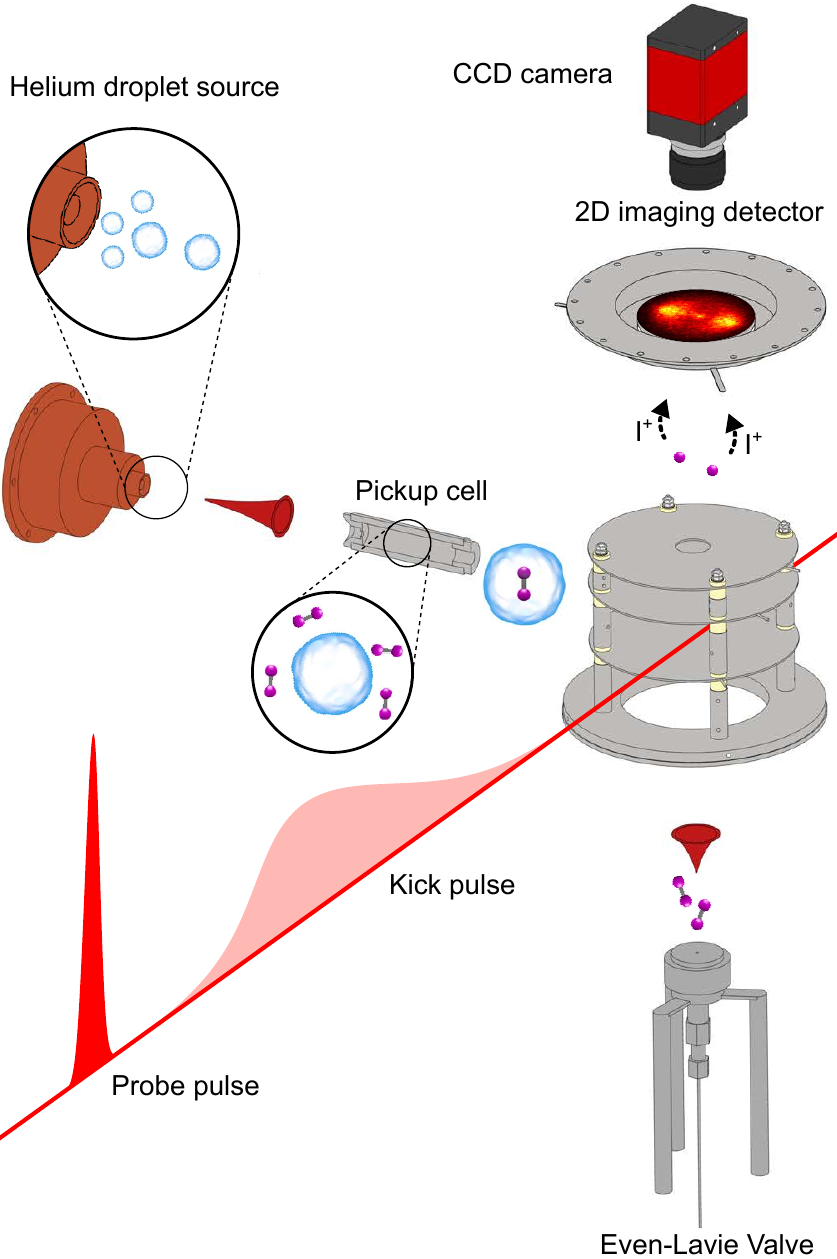}
  \caption{ {\bf Schematic of the experimental setup}. Schematic diagram showing the experimental setup used for the non-adiabatic alignment of iodine molecules both solvated inside helium droplets and isolated in a supersonic beam. Depicted from left to right are the continuous helium droplet source, the pickup cell and the 2D imaging spectrometer. Below the spectrometer sits the Even-Lavie pulsed valve used for the isolated molecule studies. The polarization state of the kick pulse (horizontal) and the delayed probe pulse (vertical) are indicated by the direction of the pulse forms sketched.}
\label{fig:exp-setup}
\end{figure}

A schematic diagram of the experimental setup used for laser-induced alignment experiments of iodine molecules, both solvated inside helium droplets and isolated in a supersonic beam, is shown in \autoref{fig:exp-setup}. Helium droplets are produced using a continuous helium droplet source with stagnation conditions of \SI{14}{K} and \SI{25}{bar}, giving  $\sim$\SI{10}{nm} diameter helium droplets~\cite{toennies_superfluid_2004}. Shortly after the exit of the continuous source the droplet beam passes through a skimmer with a \SI{1}{mm} diameter opening and enters a pickup cell containing iodine vapor. The partial pressure of the iodine vapor was kept sufficiently low to ensure the pickup of at most a single iodine molecule. Hereafter the doped droplets pass through a liquid nitrogen trap that captures the majority of the effusive iodine molecules that are not picked up by the droplets. In order to further reduce the contribution from effusive molecules the doped droplets pass through a second skimmer with a \SI{2}{mm} diameter opening followed by a second liquid nitrogen trap. Finally, the doped droplets enter the interaction region of the target chamber. In this region, the doped helium droplet beam is crossed perpendicularly by two collinear \SI{800}{nm} pulsed laser beams. The doped droplets are first irradiated with a linearly polarized kick pulse that is used to induce alignment. For the measurements up to $F_\text{kick}$ = 5.0 J/cm$^2$ the duration of the kick pulse is \SI{450}{fs}. At this duration the fluence cannot be increased further because the intensity becomes so high that the iodine molecules starts to be ionized by the kick pulse alone. The measurements with $F_\text{kick}$ = 6.4, 7.4 and 8.7 J/cm$^2$ are, therefore, recorded with a kick pulse duration of \SI{1300}{fs}. This is still much shorter than the rotational time of the iodine molecule (446 ps in gas phase) and thus keeps the experiment in the strictly non-adiabatic limit of alignment~\cite{stapelfeldt_colloquium:_2003,seideman_nonadiabatic_2005}. For consistency, we recorded the alignment experiment at $F_\text{kick}$ = 5.0 J/cm$^2$ with the \SI{1300}{fs} kick pulse and got results essentially identical to those recorded with the \SI{450}{fs} kick pulse (Fig. 1f in the main text).

After the kick pulse the molecules are Coulomb exploded by a delayed, intense probe pulse (40 fs, $\SI{3.7e14}{W/cm^2}$), which produces \ce{I^+} or \ce{IHe^+} ion fragments. The recoil directions of either ion species are given by the angular distribution of the molecular axes at the instant of the probe pulse. For the measurements reported here the \ce{IHe^+} signal was chosen as the observable because these ions can only be produced from molecules inside He droplet ~\cite{pentlehner_impulsive_2013,pentlehner_laser-induced_2013}. In the case of \ce{I^+} ions there is a contribution from those iodine molecules that manage to effuse from the pick-up cell to the interaction region in the target chamber. This contribution is, however, at most a few percent and recording of \ce{I^+} images could, therefore, also have been carried out and should give the same rotational dynamics as that obtained from the \ce{IHe^+} images.

By detecting the emission directions of the \ce{IHe^+} ions with a 2D imaging detector at many different kick-probe delays, $t$, the time-dependent degree of alignment, $\langle \cos^2 \theta_{2\mathrm{D}} \rangle$, can be determined, where $\theta_{2\mathrm{D}}$ is the angle between the kick pulse polarization and the projection of an \ce{IHe^+} ion velocity vector on the detector. The experimental setup is equipped with a pulsed Even-Lavie valve located beneath the target chamber and allows for a molecular beam of isolated iodine molecules to be sent into the interaction region. The alignment dynamics for isolated molecules was recorded under the same laser conditions as those used for the helium droplet experiments. Here \ce{I^+} ions were used as observables.

\section{Path integral Monte Carlo}

Quantum many-body systems of $N$ particles in equilibrium can be mapped to
a classical system of polymer chains~\cite{feynmanhibbs,chandlerJCP81}.
The path integral Monte Carlo (PIMC) method exploits this isomorphism.
For Bose systems like $^4$He droplets, finite-temperature results obtained by PIMC can
be considered virtually exact, given sufficient simulation time.

The PIMC method calculates equilibrium properties in the canonical
\cite{ceperley95} or grand canonical~\cite{boninsegniPRE06} ensemble.  In the
present work we use the canonical ensemble, thus
expectation values of an operator $\hat A$ are obtained as
$\langle \hat A\rangle={1\over Z}{\rm Tr}[e^{-\beta\hat H} A]$,
where $Z={\rm Tr}[e^{-\beta\hat H}]$ is the partition function, $\beta=1/k_\text{B}T$,
and $\hat H$ is the many-body Hamiltonian. In our case of a linear molecule
with bare rotational constant $B$ and mass $M$ and $N$ $^4$He atoms of mass $m$
$\hat H$ is
$$
\hat H = -{\hbar^2\over 2M}\nabla_0^2+B\hat L^2-{\hbar^2\over 2m}\sum_i\nabla_i^2
+\sum_i u(\qr_0,\qr_i,\Omega)
+\sum_{i<j} v(|\qr_i-\qr_j|)
$$
$\hat L$ is the angular momentum operator, $\Omega$ are the two Euler angles of a linear
molecule, $\qr_0$ is the center of mass coordinate of the molecule and $\qr_i$ are the coordinates
of the $^4$He atoms.  The interactions are modeled as pair-wise interactions between
$^4$He atoms, $v$, and between $^4$He atoms and the molecule, $u$.  We use the
potential by Aziz et al.~\cite{aziz87} for $v$ and the \textit{ab initio} potential
by Garcia-Gutierrez et al.~\cite{garcia-gutierrez_intermolecular_2009} for $u$.  The latter depends not only on the distance $|\qr_0-\qr_i|$ between
$^4$He atom and molecule, but also on the angle $\theta_i$ between the distance
vector $\qr_0-\qr_i$ and the axis of the molecule defined by $\Omega$.  It is the
dependence on $\theta_i$ which leads to the coupling of the rotational dynamics of
the molecule to the helium droplet.
We neglect the vibrational degree of freedom of the molecule, which for a diatomic
molecule like $I_2$ is the distance between the two iodine atoms.
The vibrational excitation energies are orders of magnitude larger than typical
rotational energies and excitation energies in helium.  Therefore, the coupling between
vibrations and helium are negligible in the study of rotational dynamics.  We assume
$I_2$ to be a rigid rotor, with the two iodine atoms separated by their
equilibrium distance of $2.666$~\AA.

For PIMC simulations, it is convenient to work in coordinate space.
Thus, for calculating expectation values $\langle\hat A\rangle$ the many-body
density matrix in configuration space, $\rho(\qR,\qR';\beta) = \langle \qR|e^{-\beta
  \hat H}|\qR'\rangle$, is sampled using
the Metropolis algorithm~\cite{metro53}.  Here $\qR$ denotes all coordinates of
the many-body system, $\qR=(\Omega,\qr_0,\qr_1,\dots,\qr_N)$.
A numerical evaluation of $\rho(\qR,\qR';\beta)$ is complicated by the fact that
in general the exponential of the many-body Hamiltonian $\hat H$ cannot be calculated.
Therefore, we split the ``imaginary time'' interval $\beta$
into small ``time steps'' $\tau=\beta/M$.  This necessitates the
introduction of new coordinates at intermediate time slices,
\begin{equation}
\label{eq_density0M}
  \rho(\qR_0,\qR_M;\beta) = \int d\qR_1\cdots\qR_{M-1}
  \rho(\qR_0,\qR_1;\tau)\cdots\rho(\qR_{M-1},\qR_M;\tau).
\end{equation}
$(\qR_0,\dots,\qR_M)$ can be regarded as a discretized path in
imaginary time.  This way each particle coordinate $\qr_i$ is replaced
by a whole path of coordinates (``beads'') $\qr_{i,j}$ where the new
index $j=0,\dots,M$ labels the discretized imaginary time.

For completing the isomorphism between a quantum system and classical
polymers of beads, we choose $\tau$ sufficiently small, such that
$\rho(\qR_0,\qR_1;\tau)$ can be approximated.  We use the pair density
approximation~\cite{ceperley95} for the He-He interaction, and the Trotter
approximation for the He-molecule interaction,
$e^{-\tau (\hat T+\hat V)}=e^{-\tau \hat V/2}e^{-\tau \hat T}e^{-\tau \hat V/2}+{\rm O}(\tau^3)$,
where $\hat T$ and $\hat V$ are the (non-commuting)
kinetic and interaction terms of the Hamiltonian $\hat H=\hat T+\hat V$, respectively.
The Trotter approximation requires
to use a relatively small time step $\tau=1/80$~K, which determines the number of
beads as $M={\beta\over\tau}$ for a given inverse temperature $\beta$.
If no off-diagonal operators such as the one-body density matrix need
to be averaged,  we can set $\qR_M=\qR_0$.  Thus all polymers are closed
loops, one for each quantum particle.

We need to account for the indistinguishability of quantum particles.
Bose statistics is implemented by symmetrization of the density matrix
\begin{equation}
\label{eq_density_sym}
  \rho_B(\qR,\qR;\beta) = \frac{1}{ N!}\sum_P \rho(\qR,P\qR;\beta)
\end{equation}
where the sum is over all permutations $P$.  As can be seen from
the right hand side of Eq.~(\ref{eq_density_sym}), the symmetrization
corresponds to reconnecting the imaginary time paths to form
larger polymers.  For a detailed review of the PIMC method for bosons
see Ref.~\cite{ceperley95}, for the application to dopants in
$^4$He clusters see Refs.~\cite{blinovJCP04,zillichJCP05}.

In addition to static quantities, PIMC allows in principle also to
calculate dynamical properties.  It is straightforward to calculate
correlation functions $\langle \hat A(t)\hat A(0)\rangle$, that are
related to measurable spectra via the fluctuation-dissipation theorem.
However, PIMC provides these correlation functions only in imaginary
time.  The analytic continuation of imaginary time data with statistical
noise due to finite sampling time to real time is an ill-posed problem.
In the present case, we are interested in the rotational spectrum of
a linear molecule, which can be obtained from
$$
  F_\ell(t) = {4\pi\over 2\ell +1}{1\over Z}\sum_m
        {\rm Tr}[Y^+_{\ell m}(t)Y_{\ell m}(0) e^{-\beta \hat H}]
$$
where $Y_{\ell m}$ are spherical harmonics.  The spectrum $S_\ell(\omega)$ of
rotational excitations $J\to J+\ell$ (where $\ell=2$ in the
case of a homo-nuclear linear rotor) can be obtained by inverting
the Laplace transform,
$F_\ell(t) = \int_{-\infty}^{\infty}d\omega e^{-t\omega}S_\ell(\omega)$.
Since the inversion is an ill-posed problem, it can only be done
approximately and only if the error bars of $F_\ell(t)$ are very
small.  Therefore, we opted to simply fit the known solution $F^0_\ell(t)$
for a free linear rotor,
$$
  F_\ell(t) = {4\pi\over 2\ell +1}{3\over Z}\sum_{\ell_1,\ell_2}
        {(2\ell_1+1)(2\ell_2+1)\over 4\pi}
        \left({{\ell_1 \atop 0} {\ell \atop 0} {\ell_2 \atop 0}}\right)^2
        e^{-(\beta-t)B\ell_2(\ell_2+1)}
        e^{-tB\ell_1(\ell_1+1)}
$$
to correlation function $F_\ell(t)$ obtained with PIMC for $I_2$ in helium
with $B$ acting as fit parameter.  This fit yields an effective
$B=B_{\rm eff}$, under the assumption that the rotational spectrum
of $I_2$ in helium is essentially that of a linear rotor with a renormalized
rotational constant.  For heavy rotors this assumption has been validated
by experiments.  Although the effective distortion constant $D_{\rm eff}$ usually increases
by orders of magnitude in helium compared to the gas phase value,
$D_{\rm eff}$ is still small and, therefore, we cannot determine it from
an improved fit to a free linear rotor with distortion constant $D$ as
second parameter.

PIMC simulations of droplets of $10^3$ or $10^4$ $^4$He atoms, as produced
in the experiments, would be very demanding.  Instead
we performed two kinds of simulations of $I_2$ in helium, that bracket the
experimental situation from both sides: $I_2$
in $^4$He clusters of $N=150$ atoms, much smaller than in experiment;
and $I_2$ in bulk $^4$He.  Bulk
simulations are realized with periodic boundary conditions.
511 $^4$He atoms and one $I_2$ are put in a
simulation box of side length $L=28.6$\AA.  $L$ is determined by the
condition that the $^4$He density in the molecule frame of reference,
$\rho(r,\theta)$, approaches the
equilibrium density of bulk $^4$He, $\rho_{eq}=0.02186\,$\AA$^{-3}$, for
large distance $r$ between $^4$He atoms and molecule.  The largest distance
compatible with periodic boundaries is $L/2$, which is large enough to
obtain a $^4$He density $\rho(r,\theta)$ that fluctuates only slightly
around $\rho_\text{eq}$.  Therefore, we believe that simulations for this size
provide a good approximation to the bulk limit.  Simulations of
$I_2$ in $^4$He cluster were done at a temperature of $T=0.31$K, typical
for $^4$He droplets in equilibrium.  For simulations of $I_2$ in bulk
helium, approximated by 511 $^4$He atoms, we doubled the temperature
to $T=0.62$K, in order to reduce the computational demands by cutting
the number of beads in half.
From the bulk simulations we obtained a ratio of $B_{\rm eff}/B=0.60$, while simulations
of I$_2$ in a cluster of $N=150$ $^4$He atoms gave a very similar value
of $B_{\rm eff}/B=0.58$. We note that for I$_2$ the ratio $B_{\rm eff}/B$ is large compared
to values found for other heavy linear rotors in helium.  This smaller relative
reduction is due to the particularly large moment of inertia $I$ of I$_2$ already
in the gas phase.  A significant {\em relative} increase of $I$, and thus significant {\em relative}
reduction of $B$ requires, therefore, a much larger effect of the helium environment
than, e.g., for the well-studied OCS molecule in helium, with a moment of inertia
more than five times smaller.

\section{The angulon theory}

\subsubsection{The angulon Hamiltonian}

The theoretical approach used here is based on the recently-developed angulon theory~\cite{schmidt_rotation_2015, schmidt_deformation_2016, LemeshkoDroplets16, LemSchmidtChapter,  MidyaPRA16, Yakaboylu17, RedchenkoCPC16, Li16}.
We start from the angulon Hamiltonian, which describes a rotating molecule coupled to a bosonic bath~\cite{schmidt_rotation_2015}:
\begin{equation}
\label{hamiltonian}
 \widehat{H} = B \hat{\mathbf{J}}^2 + \sum\limits_{k\lambda\mu} \omega_k \hat{b}^\dag_{k\lambda\mu} \hat{b}_{k\lambda\mu}
 + \sum\limits_{k\lambda\mu} U_\lambda(k) ~ [Y^*_{\lambda\mu}(\hat{\theta},\hat{\phi})\hat{b}^\dag_{k\lambda\mu}+Y_{\lambda\mu}(\hat{\theta},\hat{\phi}) \hat{b}_{k\lambda\mu}],
\end{equation}
where we used the notation  $\sum_k \equiv \int dk$, and set $\hbar \equiv 1$. The first term of Eq.~\eqref{hamiltonian} corresponds to the rotational kinetic energy of a linear-rotor molecule, with $\hat{\mathbf{J}}$ the angular momentum operator and $B = 1/(2I)$ the molecular rotational constant, where $I$ is the molecular moment of inertia. The bare eigenstates of the molecular impurity are  given by the $(2L+1)$-fold degenerate states, $ \vert L,M\rangle$, with energies $E_L = B L(L + 1)$. Here $L$ is the angular momentum quantum number, and $M$ is its projection on the laboratory-frame $z$-axis.

The second term of Eq.~\eqref{hamiltonian} gives the kinetic energy of the superfluid excitations, such as phonons and rotons, whose spectrum is described by the dispersion relation $\omega_k$. Here, the operators $\hat{b}^\dag_{k\lambda\mu}$ ($\hat{b}_{k\lambda\mu}$) are creating (annihilating) a superfluid excitation with linear momentum $k=|\mathbf{k}|$, the angular momentum $\lambda$, and its projection, $\mu$, onto the $z$-axis. These operators can be obtained using the spherical-harmonic expansion of the usual creation/annihilation operators in Cartesian space,  $\bed_\veck$ and $\be_\veck$, see Refs.~\cite{schmidt_rotation_2015, schmidt_deformation_2016, LemSchmidtChapter} for details.

The last term of the Hamiltonian~\eqref{hamiltonian} describes the angular-momentum exchange between the molecular impurity and the superfluid, where the coupling constants $U_\lambda(k)$ are proportional to the Legendre moments of the molecule--Helium potential energy surface   in  Fourier space. Here $Y_{\lambda\mu}(\hat{\theta},\hat{\phi})$ are spherical harmonics~\cite{Varshalovich}, which depend on the molecular angle operators in the laboratory frame, $(\hat{\theta},\hat{\phi})$. This type of coupling, explicitly dependent on the three-dimensional impurity orientation, makes Eq.~\eqref{hamiltonian} substantially different from other impurity problems such as, e.g., the Bose-polaron~\cite{Devreese15} or the spin-boson~\cite{LeggettRMP87} models.

Originally, the Hamiltonian~\eqref{hamiltonian} was  derived to describe an ultracold molecule interacting with a dilute BEC, where the coupling constants $U_\lambda(k)$ assume a simple analytic form~\cite{schmidt_rotation_2015,LemSchmidtChapter}. Helium, on the other hand, represents a dense, strongly-interacting superfluid, which makes it quite challenging to derive the coupling constants from first principles.

However, by analogy with effective field theories of nuclear~\cite{MachleidtPR11} and condensed matter~\cite{FradkinFTCM} physics, the angulon Hamiltonian~\eqref{hamiltonian} can be approached from a phenomenological perspective, where the effective low-energy constants are extracted from experiment or \textit{ab initio} calculations. As an example, recently it was shown that the effective rotational constants of 25 different molecules in superfluid helium can be obtained from the angulon theory in good agreement with experiment, based on only two phenomenological parameters~\cite{LemeshkoDroplets16}. Here we pursue a similar approach to calculate the dynamical properties of the I$_2$ molecule in helium.

\vspace{0.2cm}
\subsubsection{Effective rotational constants}

Interactions of a heavy molecule, such as I$_2$, with helium can be most naturally understood if one rewrites the Hamiltonian~\eqref{hamiltonian} in the rotating molecular frame~\cite{LemeshkoDroplets16}. This is achieved by applying a canonical transformation recently introduced by Schmidt and Lemeshko~\cite{schmidt_deformation_2016}:
\begin{equation}
\label{Transformation}
	 \hat{S} = e^{- i \hat\phi \otimes \hat \Lambda_z} e^{- i \hat\theta  \otimes \hat\Lambda_y} e^{- i \hat\gamma  \otimes\hat \Lambda_z}. \\
\end{equation}
Here $(\hat\phi, \hat\theta, \hat\gamma)$ are the angle operators which act in the  Hilbert space of the molecular rotor, and
\begin{equation}
\label{Lambda}
	 \hat {\mathbf\Lambda}=\sum_{k\lambda\mu\nu}\bed_{k\lambda\mu}\boldsymbol\sigma^{\lambda}_{\mu\nu}\be_{k\lambda \nu}
\end{equation}
is the total angular momentum operator of the superfluid excitations, acting in their corresponding Hilbert space. The angular momentum matrices, $\boldsymbol\sigma^{\lambda} \equiv \{\sigma^{\lambda}_{- 1}, \sigma^{\lambda}_{0}, \sigma^{\lambda}_{+1} \}$,  fulfill  the $SO(3)$ algebra in the representation of angular momentum $\lambda$.
Thus, the  transformation of Eq.~\eqref{Transformation} transfers the superfluid degrees of freedom into the rotating molecular frame.

The transformed Hamiltonian assumes the following form:
\begin{equation}
\label{transH}
\hat{\mathcal{H}} \equiv \hat S^{-1} \hat H \hat S= B (\hat{\mathbf{L}} - \hat{\mathbf{\Lambda}})^2   + \sum_{k\lambda\mu}\omega_k \bed_{k\lambda\mu}\be_{k\lambda\mu} + \sum_{k\lambda} \tilde U_\lambda(k) \left[\bed_{k\lambda0}+\be_{k\lambda0}\right]
\end{equation}
where $\tilde{U}_\lambda(k) = \sqrt{(2 \lambda +1)/(4 \pi)} U_\lambda(k)$. The operator $\hat{\mathbf{L}} \equiv \hat{\mathbf{J}} + \hat{\mathbf{\Lambda}}$ is the \textit{total} angular momentum operator, which acts in the molecular Hilbert space. The components of $\hat{\mathbf{L}}$ define projections of total angular momentum in the rotating molecular frame and, therefore, obey anomalous commutation relations~\cite{LevebvreBrionField2, BiedenharnAngMom}. In the absence of external fields, total angular momentum $\hat{\mathbf{L}}$ is conserved, which allows to solve the problem for each value of $L$ separately.

Another advantage of the transformed Hamiltonian~\eqref{transH} is that it can be diagonalized exactly in the limit of a slowly rotating molecule, $B \to 0$. There, for each total angular momentum state, $\ket{L M}$, the ground state is given by:
\begin{equation}
\label{transHground}
	\ket{\psi_{LM}} =  e^{  \sum_{k \lambda}  \frac{\tilde{U}_\lambda (k)}{\omega_{k}} \left( \be_{k \lambda 0} - \bed_{k \lambda 0}  \right)} \ket{0} \ket{LM}.
\end{equation}

Note that the bosonic coherent state of Eq.~\eqref{transHground} involves an infinite number of superfluid excitations and, therefore, describes a collective anizotropic displacement of helium atoms. Such a deformation can be thought of as a microscopic formulation of the `nonsuperfluid helium shell' which rotates along with the molecule~\cite{grebenev_rotational_2000, toennies_superfluid_2004}.

It is important to note that the angulon theory based on the transformed Hamiltonian, Eq.~\eqref{transH}, provides a simple physical explanation for renormalization of molecular rotational constants in superfluid helium. The rotational energy of the molecular impurity is defined by the first term of Eq.~\eqref{transH}, while the rest of terms ultimately determine how many phonons are excited due to the molecule-helium interactions. In the absence of helium, the total angular momentum is given by that of a free molecule, $ \hat{\mathbf{L}} \equiv \hat{\mathbf{J}}$. In the presence of helium, $ \hat{\mathbf{L}}$ is still a conserved quantity, however, the stronger  the molecule-helium interactions the larger is the angular momentum of the superfluid, $\hat{\mathbf{\Lambda}}$. Thus, for a given \textit{total} angular momentum $L$, the rotational energy is lower in the presence of helium  ($\hat {\mathbf\Lambda} \neq 0$) compared to a gas-phase molecule ($\hat {\mathbf\Lambda} = 0$), which leads to renormalization of the molecular rotational constant.

In our approach, we calculate the amount of the angular momentum transferred to the superfluid for the state of Eq.~\eqref{transHground}:
\begin{equation}
\label{ExpLambda}
	  \langle  \hat{\mathbf{\Lambda}}^2 \rangle \equiv \bra{\psi_{LM}} \hat{\mathbf{\Lambda}}^2 \ket{\psi_{LM}} = \sum_{k \lambda} \lambda(\lambda+1) \frac{\tilde{U}^2_\lambda (k)}{\omega^2_{k}},
\end{equation}
and replace the boson angular momentum operator in Eq.~\eqref{transH} by its expectation value, $\hat{\mathbf{\Lambda}} \to \langle \Lambda \rangle \equiv \langle  \hat{\mathbf{\Lambda}}^2 \rangle^{1/2}$. Then, assuming that $\hat{\mathbf{\Lambda}}$ points along the total angular momentum, $\hat{\mathbf{\Lambda}} \sim \hat{\mathbf{L}}$, we can evaluate the effective rotational constant as:
\begin{equation}
\label{Bstar}
	 B_\text{eff} = B \left( 1-  \beta  \right)^2,
\end{equation}
where
\begin{equation}
\label{beta}
	\beta= \left( \frac{1}{2}\sum_{k,\lambda}  \lambda(\lambda+1) \frac{\tilde U_\lambda(k)^2}{\omega^2_{k}} \right)^{1/2}.
\end{equation}
Within our approach, we treat $\beta$ as a phenomenological parameter and set it to $\beta=0.23$, which reproduces the results of quantum Monte Carlo calculations giving $B_\text{eff} = 0.6 B$.  In addition, Eq.~\eqref{beta} enables us to calculate the $\alpha_2$ parameter as discussed below.

\vspace{0.2cm}

\subsubsection{Dynamics of I$_2$ in helium}

We perform the calculations of the time evolution in the laboratory frame, as given by the Hamiltonian~\eqref{hamiltonian}, with $B$ replaced by  $B_\text{eff} = 0.6 \, B_0$, as discussed above. Since the pulse is very short, $\tau \ll B^{-1}, U_\lambda(k)^{-1}, \omega_k^{-1}$, the state after the pulse can be found within the impulsive approximation~\cite{henriksen_molecular_1999}. To this end, we assume that the interaction with the laser pulse is described by the potential
\begin{equation}
\hat V(t) = - \eta \, \delta(t) \cos^2{\hat{\theta}}. \label{eq:laser}
\end{equation}
Here, $\eta$ is a dimensionless parameter related to the fluence of the kick pulse. The laser fluences presented in  Fig.~1(a)--(e) correspond to the values of $\eta = 2.6, 5.2, 13, 26$, and $39$, respectively.
The subsequent time evolution of the wave function is given by:
\begin{equation}
|\psi(t)\ran = e^{-i \hat{H} t} e^{\eta \cos^2 \hat{\theta}} |\psi_0 \ran. \label{eq:evo1}
\end{equation}
Before the laser pulse, the He-dressed molecular states in the laboratory frame are obtained by applying an inverse transformation of Eq.~\eqref{Transformation} to Eq.~\eqref{transHground}, which results in:
\begin{equation}
\label{eq:psi0ev}
|\psi_0 \ran = e^{  \sum_{k \lambda \mu}  \frac{\tilde{U}_\lambda (k)}{\omega_{k}} \left( \be_{k \lambda \mu} Y_{\lambda \mu}(\hat{\Omega})  - \bed_{k \lambda \mu} Y_{\lambda \mu}^*(\hat{\Omega})  \right)} |j_0, m_0\ran_{\rm mol} \, \otimes |0 \ran_{\rm b},
\end{equation}
where $\hat \Omega \equiv (\hat \theta, \hat \phi)$ and $|0 \ran_{\rm b}$ is the bosonic vacuum state. Since after the pulse the wave function represents a superposition of rotational energy levels, the time-evolution of the angulon state is given by:
\begin{equation}
|\psi(t)\ran = e^{-i \hat{H} t} \sum_j c_j e^{  \sum_{k \lambda \mu}  \frac{\tilde{U}_\lambda (k)}{\omega_{k}} \left( \be_{k \lambda \mu} Y_{\lambda \mu}(\hat{\Omega})  - \bed_{k \lambda \mu} Y_{\lambda \mu}^*(\hat{\Omega})  \right)} |j, m_0\ran_{\rm mol} \otimes |0 \ran_{\rm b},  \label{eq:tdep}
\end{equation}
where the coefficients $c_j$ depend on $j_0$, $m_0$, and $\eta$.
For the  wave function \eqref{eq:tdep}, we calculate the evolution of the molecular alignment as
\begin{equation}
\lan \cos^2{\theta_{\rm 2D}}\ran (t) \equiv \lan \psi(t) | \cos^2{\hat{\theta}_{\rm 2D}} | \psi(t) \ran,
\end{equation}
where $\cos^2 \hat \theta_{\rm 2D} \equiv \cos^2{\hat \theta}/(\cos^2{\hat \theta} + \sin^2{\hat \theta}\sin^2{\hat \phi})$ is a two-dimensional projection of the three-dimensional alignment cosine, $\cos^2 \hat \theta$, as measured in the experiments.

Performing the time evolution of the considered wave function using Eqs.~(\ref{hamiltonian}) and (\ref{eq:tdep}) is an involved many-particle problem because the interaction term of the Hamiltonian does not commute with the rest of the terms. First of all, we note that the boson kinetic energy can be eliminated by rewriting the Hamiltonian in the rotating frame, which corresponds to the replacements $\bed_{k \lambda \mu} \to \bed_{k \lambda \mu} e^{- i \omega_k t}$, $\be_{k \lambda \mu} \to \be_{k \lambda \mu} e^{i \omega_k t}$. Since in the final expression for the alignment cosine the oscillating exponents cancel with their corresponding complex conjugates, the boson kinetic energy contributes only through the initial state, Eq.~\eqref{eq:psi0ev}. In order to account for the rest of the Hamiltonian, we apply an expansion of the Suzuki-Trotter type:
\begin{equation}
\label{eq:ST}
e^{-i t \hat{H}} \approx e^{-i t B \mathbf{\hat{J}^2} / 2} e^{-i t \sum_{k \lambda \mu} U_\lambda(k)  \left[ Y^\ast_{\lambda \mu} (\hat \theta,\hat \phi) \bed_{k \lambda \mu}+ Y_{\lambda \mu} (\hat \theta,\hat \phi) \be_{k \lambda \mu} \right]} e^{-i t B \mathbf{\hat{J}^2} / 2},
\end{equation}
which becomes exact in the limit of small $t$. Furthermore, in order to decrease the number of free parameters of the model, we take into account only the leading anisotropic term, $\lambda=2$, of the He--I$_2$ PES~\cite{garcia-gutierrez_intermolecular_2009, LemeshkoDroplets16}. The resulting alignment cosine can be derived in closed form:
\begin{align}
\lan \cos^2{\hat{\theta}_{\rm 2D}}\ran (t) = \sum_{j, j', j_1, j_2, m} c^*_{j'} c_{j} e^{i t (E_{j_2} + E_{j'} - E_{j_1} - E_{j}) / 2} \lan j_2 m| \cos^2{\hat{\theta}_{2D}} |j_1 m\ran \nonumber \\
\times \int d\Omega_1 \int d\Omega_2 \: Y^*_{j' m_0}(\Omega_2) Y_{j_2 m}(\Omega_2) Y_{j m_0}(\Omega_1) Y^*_{j_1 m}(\Omega_1) e^{(\alpha_1 t^2 + \alpha_2)  \left[ \frac{4 \pi}{5}  \sum_{\mu}   Y_{2, \mu}(\Omega_2) Y^*_{2, \mu}(\Omega_1) - 1 \right]},\label{eq:main}
\end{align}
where $\int d\Omega \equiv \int \sin{\theta} \, d\theta \, d\phi$  and we have defined
\begin{equation}
\alpha_1 \equiv \sum_{k} U^2_2(k), \qquad
\alpha_2 \equiv \sum_{k} \frac{U^2_2(k)}{\omega^2_k}.
\end{equation}
The $\alpha_1$ parameter determines the decay rate of the alignment. We take $\alpha_1 = 10$ (in units of $1/B_\text{eff}^{2}$), which reproduces the decay rate observed in experiment for \mbox{$F_\text{kick}$ = 1.2 J/cm$^2$}. The $\alpha_2$ parameter can be determined using Eq.(\ref{ExpLambda}) and the value of $B_\text{eff} = 0.6 B_0$ and turns out to be very small $\alpha_2 \approx 0.04$.
Previous experiments on microwave spectroscopy of molecules in helium droplets have shown a  broadening of the molecular rotational lines, which amounts to  $\sim100$~MHz in the case of OCS~\cite{lehnig_rotational_2009}.   The line shapes are attributed to inhomogeneous broadening, and at least for CO in helium this was confirmed by calculations~\cite{zillichJCP08}. In order to account for this effect, we introduce a Gaussian broadening of $B_\text{eff}$ values with a standard deviation $\gamma = 0.05~B_\text{eff}$, which corresponds to half width at half maximum of $66$~MHz for I$_2$. Explicitly, we use the energy levels $E_j \equiv (B_\text{eff} + \gamma  x) j(j+1)$, where $x$ is drawn from a normal distribution and we integrate the result for $\lan \cos^2{\theta_{\rm 2D}}\ran (t)$ over $x$.

In addition, we calculate the survival probability of the initial state as:
\begin{equation}
S(t) = \left|\sum_j |c_j|^2 e^{-i t E_j}\right|^2 e^{-\alpha_1 t^2}.
\end{equation}

\vspace{0.2cm}
\subsubsection{Ensemble averaging: temperature and symmetry}

Since the temperature of He droplets, $T=0.38 \; {\rm K}$, is larger than the rotational constant $B_\text{eff} = 0.032 \; {\rm K}$, one needs to account for the thermal distribution of population over several rotational states, as given by the Boltzmann distribution:
\begin{equation}
P_j = Z^{-1} e^{- B_\text{eff} j (j+1) / (k_B T)}.
\end{equation}
where $k_B$ is the Boltzmann constant and $Z \equiv \sum_j P_j$ is the partition function. For the temperature stated above, it is sufficient to truncate the sum at $j=6$, which corresponds to $28$ lowest $\ket{jm}$-levels.

In addition to the thermal distribution, we account for the $21:15$ ortho-to-para ratio of I$_2$. This corresponds to an averaging over molecules in even and odd rotational states:
\begin{align}
P_j^{\rm even} & = Z_{\rm even}^{-1} e^{- B_\text{eff} j (j+1) / (k_B T)}, \\
P_j^{\rm odd} & = Z_{\rm odd}^{-1} e^{- B_\text{eff} [ j (j+1) - 2] / (k_B T)},
\end{align}
where the reference energy for odd states is $2 B_\text{eff}$ -- the rotational energy of the lowest odd state with $j=1$. The corresponding partition functions are given by $Z_{\rm even (odd)} \equiv \sum_{j \in {\rm even (odd)}} P_j^{\rm even (odd)}$. The thermally-averaged result for the alignment, $a(t)$, of the odd and even states is given by:
\begin{equation}
\lan \cos^2{\theta_{\rm 2D}}\ran_{\rm even(odd)}(t) = \sum_{j_0 \in {\rm even(odd)}, m_0} \lan \psi(t) | \cos^2{\theta_{({\rm 2D})}} | \psi(t) \ran^{(j_0, m_0)} \times P_j^{\rm even (odd)},
\end{equation}
where the ${(j_0, m_0)}$ superscript denotes alignment with the starting $|j_0, m_0\ran$ state of the molecule. Finally, the ensemble-averaged result is obtained as
\begin{equation}
\label{eq:aAver}
\lan \cos^2{\theta_{\rm 2D}}\ran (t) = \frac{15 \times \lan \cos^2{\theta_{\rm 2D}}\ran_{\rm even}(t) + 21 \times \lan \cos^2{\theta_{\rm 2D}}\ran_{\rm odd}(t)}{15 + 21}.
\end{equation}
Eq.~\eqref{eq:aAver} was the one we used to compare the theory  to experiment.

\end{document}